\documentclass[12pt]{nature}
\usepackage{aas_macros}
\usepackage{amssymb}
\usepackage{amsmath}
 \usepackage{graphicx}  
\usepackage[svgnames]{xcolor}
\usepackage{lineno}
\usepackage{placeins}

\makeatletter

\newcommand{\Rmnum}[1]{\expandafter\@slowromancap\romannumeral #1@}
\makeatother

\newcommand  \HII{\,H\,{\footnotesize II}}
\newcommand  \HI{\,H\,{\footnotesize I}}
\newcommand  \m{\mathrm}

\title{The star-forming complex LMC-N79 as a future rival to 30 Doradus}
\author{Bram B. Ochsendorf$^{1,\ast}$, Hans Zinnecker$^{2,3}$, Omnarayani Nayak$^{1}$, John Bally$^{4}$, Margaret Meixner$^{1,5}$, Olivia C. Jones$^{5}$, Remy Indebetouw$^{6,7}$ \& Mubdi Rahman$^{1}$}

\begin{document}
\maketitle

\begin{affiliations}
 \item Department of Physics and Astronomy, The Johns Hopkins University, 3400 North Charles Street, Baltimore, MD 21218, USA, bochsen1@jhu.edu
 \item Deutsches SOFIA Institut (DSI), University of Stuttgart, Pfaffenwaldring 29, D-70569, Germany
 \item Universidad Autonoma de Chile, Santiago de Chile, Chile
 \item Astrophysical and Planetary Sciences Department, University of Colorado, UCB 389 Boulder, Colorado 80309, USA
 \item Space Telescope Science Institute, 3700 San Martin Drive, Baltimore, MD 21218, USA
 \item Department of Astronomy, University of Virginia, PO 400325, Charlottesville, VA 22904, USA
 \item National Radio Astronomy Observatory, 520 Edgemont Rd, Charlottesville, VA 22903, USA
\end{affiliations}

\begin{abstract}

Within the early Universe, `extreme' star formation may have been the norm rather than the exception\cite{barger_1998,turner_2009}. Super Star Clusters (SSCs; $M_\star$\,$\gtrsim$\,10$^5$ M$_\odot$) are thought to be the modern-day analogs of globular clusters, relics of a cosmic time ($z$\,$\gtrsim$\,2) when the Universe was filled with vigorously star-forming systems\cite{kruijssen_2015}. The giant HII region 30 Doradus in the Large Magellanic Cloud (LMC) is often regarded as a benchmark for studies of extreme star formation\cite{walborn_1991}. Here, we report the discovery of a massive embedded star forming complex spanning $\sim$\,500 pc in the unexplored southwest region of the LMC, which manifests itself as a younger, embedded twin of 30 Doradus. Previously known as N79, this region has a star formation efficiency exceeding that of 30 Doradus by a factor of $\sim$\,2 as measured over the past $\lesssim$\,0.5 Myr. Moreover, at the heart of N79 lies the most luminous infrared (IR) compact source discovered with large-scale IR surveys of the LMC and Milky Way, possibly a precursor to the central SSC of 30 Doradus, R136. The discovery of a nearby candidate SSC may provide invaluable information to understand how extreme star formation proceeds in the current and high-redshift Universe.
\end{abstract}
\vspace{8mm}

The LMC is the prototypical `Barred Magellanic Spiral', a population of galaxies with an asymmetric, sometimes off-centered stellar bar, a single spiral arm, and often a large star forming complex at one end of the bar\cite{de_vaucouleurs_1972}. More recently, evidence of {\em multiple} arm-like features extending from the outer disc of the LMC were obtained with high-resolution \HI\ maps\cite{staveley-smith_2003}, thought to originate from tidal interactions with both the Galaxy and Small Magellanic Cloud (SMC)\cite{bekki_2007} (Fig. 1a). In particular, at heliocentric velocities of $\sim$\,255\,-\,270 km\,s$^{-1}$ (Fig. 1b) the LMC resembles a barred spiral galaxy with two prominent, opposing arms extending from the eastern (Arm E) and western (Arm W) part of the \HI\ disk.

Arm E culminates in the `south-eastern \HI\ overdensity' located at the leading edge of the LMC's motion through the Galactic Halo\cite{de_boer_1998}. At the tip of the south-eastern \HI\ overdensity and Arm E lies 30 Doradus, harboring the largest \HII\ region in the Local Group. Ionized gas traces massive star populations with a median age of $\sim$\,4 Myr\cite{murray_2011}. In addition, with the use of sensitive, galaxy-wide IR surveys of the LMC with {\em Spitzer} and {\em Herschel}, some $\sim$\,3500 Young Stellar Object (YSO) candidates have been identified through color-magnitude cuts, morphological inspection, and spectroscopic follow-up observations (see Methods). From these YSO candidates, we have identified a subset of {\em massive} YSOs (MYSOs) that are well characterized by YSO models. This MYSO catalogue is tested to be complete for massive ($M$\,$\textgreater$\,8\,$M_\m{\odot}$), young ($\lesssim$\,0.5 Myr) objects, and this data offers a unique snapshot of the {\em most recent} massive star formation activity of the LMC\cite{ochsendorf_2016}. 

\begin{figure*}[t]
\centering
\includegraphics[width=16.5cm]{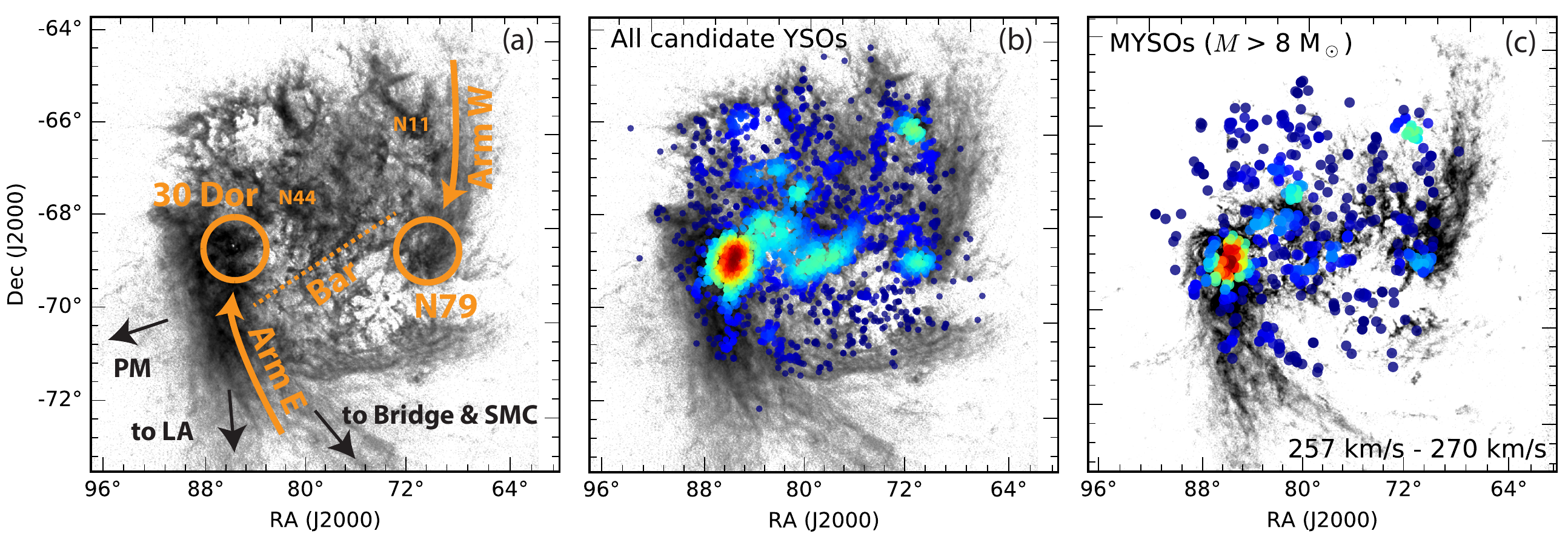} 
\caption{{\bf Large-scale structure of the LMC.} {\bf (a)}: \HI\ map of the LMC. Black and orange markings highlight locations of prominent features in the LMC and the larger-scale Magellanic complex, such as the assymetrical, off-centered optical stellar bar (see Fig. \ref{fig:overview}), and tidal arms E and W. Arm E culminates in the south-eastern \HI\ overdensity, splits at the LMC tidal radius, and subsequently leads to the Magellanic Bridge/SMC and the Leading Arm (LA). The proper motion (PM) of the LMC through the Galaxy is also indicated. In orange we highlight the rotation of the LMC disk\cite{van_der_marel_2014} and the locations of several prominent star forming regions. {\bf (b)}: Same as {\bf (a)}, but here we overlay $\sim$\,3500 YSO candidates discovered in various galaxy-wide surveys of the LMC (see Methods). The color code reflects the surface density of YSOs, with red indicating the highest local surface density or intrinsic clustering of these sources. Clearly discerned are the stellar bar and several star forming regions: 30 Doradus, N11, N44, and the relatively unknown complex N79. {\bf (c)}:  Same \HI\ map, but only showing the velocity range 257\,-\,270 km s$^{-1}$ to accentuate the `barred spiral' appearance of the LMC. Here, we overlay the subset of massive YSOs (MYSOs; $M$\,$\textgreater$\,8 $M_\odot$). Luminous, embedded, and extremely young, these MYSOs offer a snapshot to the massive star formation activity of the LMC averaged over the past $\lesssim$\,0.5 Myr\cite{ochsendorf_2016}.}
\label{fig:hi}
\end{figure*}

In Fig. 1b, we overplot the \HI\ map of the LMC with the location and clustering of YSO	candidates found across the galaxy. Several obvious clusterings stand out: the stellar bar (which likely contains many false positives; see Methods), and the well-known star forming regions 30 Doradus, N11 and N44. In addition, Fig. 1b reveals a star forming complex in the relatively unexplored southwest region of the LMC, which coincides with the N79 \HII\ region\cite{henize_1956}. Figure 1c plots the subsample of MYSOs over the \HI\ gas at a velocity range 257\,-\,270 km s$^{-1}$, which highlights the tidal arms of the LMC. Most interestingly, both 30 Doradus and N79 are perched on the leading edges or `tips' of the opposite tidal arms E and W, respectively. 

\begin{figure*}[t]
\centering
\includegraphics[width=16.5cm]{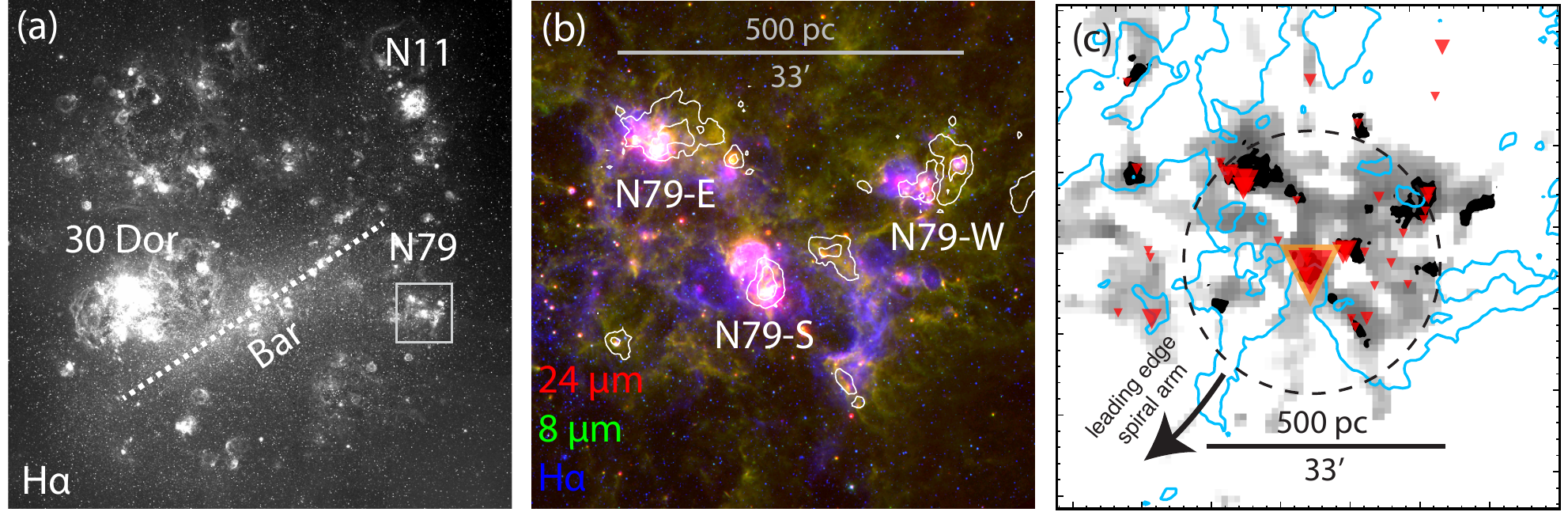} 
\caption{{\bf Dissecting N79.} {\bf (a)}: H$\alpha$ image of the LMC. Highlighted are N79 (grey box), 30 Doradus, N11, and the stellar bar. {\bf (b)}: Blow-up of the N79 region in a three-color image showing H$\alpha$ ({\em blue}), Spitzer/IRAC 8$\mu$m ({\em green}), and Spitzer/MIPS 24$\mu$m ({\em red}). White contours show CO clouds from the MAGMA survey, where we identify three main CO sub-complexes: N79-South (S), N79-East (E), and N79-West (W) {\bf (c)}: 
The CO-based ({\em filled black contour}) and dust-based ({\em grayscale}) molecular mass in N79, overplotted with the location of MYSOs ({\em inverted red triangles}; size reflects source luminosity). Also plotted is the \HI\ gas ({\em blue contours}). While CO peaks in distinct regions, harboring apparent clusterings of MYSOs, the entire complex is bridged through molecular gas as traced by dust, which is sensitive to the extended, more diffuse envelopes of GMCs\cite{ochsendorf_2017}.}
\label{fig:overview}
\end{figure*}

The N79 \HII\ region pales in comparison to optically bright star forming regions such as N11 or 30 Doradus (Fig. 2a). Hence, N79 has not been the subject of any prior high-resolution study. However, our IR observations trace the younger, more embedded phase of massive star formation and unveil that the N79 region is a highly efficient star forming engine, exceeding the star formation efficiency of 30 Doradus and N11 by a factor of $\sim$2.0\,-\,2.5 (see below). 

{\em Spitzer} and {\em Herschel} dissect the structure of the complex, spanning roughly 500 pc, and harboring three main CO complexes: N79-South, N79-East, and N79-West (Fig. 2b). The $^{12}$CO\,(1-0) tracer is known to probe a limited range in volume densities of molecular gas because of critical density, depletion, opacity, and photo-chemical effects. In addition, at the reduced metallicity of the LMC, a significant part of H$_2$ may be in a `CO-dark' phase\cite{madden_1997}. By combining far-infrared dust emission and \HI\ one can circumvent these limitations and estimate the H$_\m{2}$ distribution\cite{jameson_2016}. The dust-based molecular material (Fig. 2c) shows that the entire N79 region consist of one single molecular structure of $\sim$\,500 pc. This unusual large size may be the result of gas accumulation and compression at the tip of Arm W. Star formation concentrates within the molecular material, with apparent clusterings in the CO-emitting clouds (Fig. 2c). 

\begin{figure*}[t]
\centering
\includegraphics[width=8.3cm]{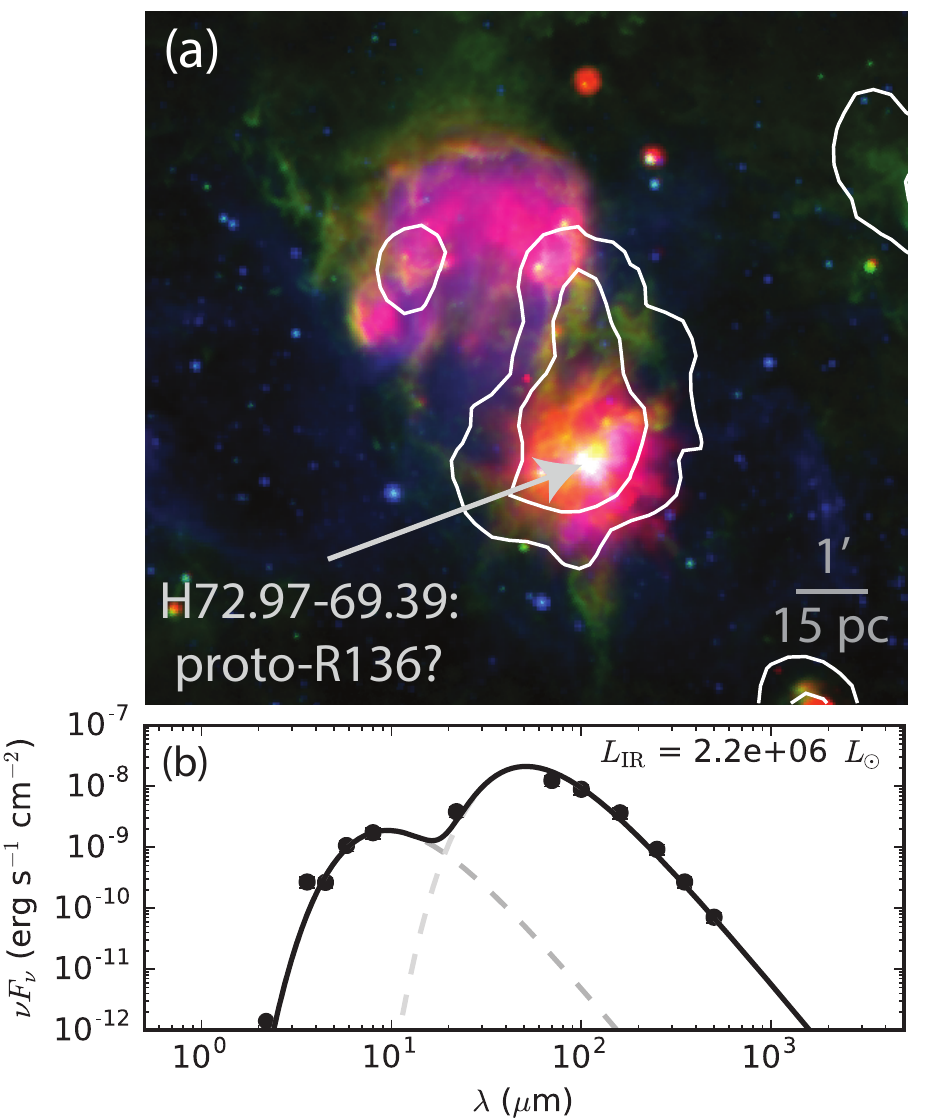} 
\caption{{\bf H72.97-69.39.} {\bf (a)}: The immediate environment of the compact luminous object at the heart of N79, H72.97-69.39, possibly a precursor to the R136 cluster in 30 Doradus (Nayak et al., in prep). {\bf (b)}: Spectral energy distribution of H72.97-69.39 compiled from various ground and space-based surveys (see Methods). A two-temperature modified blackbody yields an infrared luminosity of $L_\m{IR}$ = 2.2\,$\times$\,10$^6$ $L_\odot$.}
\label{fig:n79}
\end{figure*}

At the heart of the large-scale N79 complex lies an extremely luminous object (Fig. 3a), which immediately draws parallels to the central cluster of 30\,Dor, R136 (Nayak et al., in prep). This source has been catalogued\cite{seale_2014} as HSOBMHERICC\_J72.971176-69.391112, but will be referred to as `H72.97-69.39'. At $L_\m{IR}$\,$\simeq$\,2.2\,$\times$\,10$^6$ $L_\odot$ (Fig. 3b), H72.97-69.39 is more luminous than any MYSO or compact \HII\ region discovered with large-scale IR surveys of the LMC\cite{seale_2014} and Milky Way\cite{mottram_2011}. This luminosity is equivalent to more than three O3V stars of $M$$\sim$\,70\,M$_\odot$\cite{mottram_2011} or a single very massive star of $\sim$\,160 M$_\odot$, using the mass-luminosity relation for upper-main sequence stars\cite{zinnecker_2007}, $L$\,$\propto$\,$M^{1.6}$ (at $M$\,$\gtrsim$\,70\,M$_\odot$).  
 
\begin{figure*}[t]
\centering
\includegraphics[width=16.5cm]{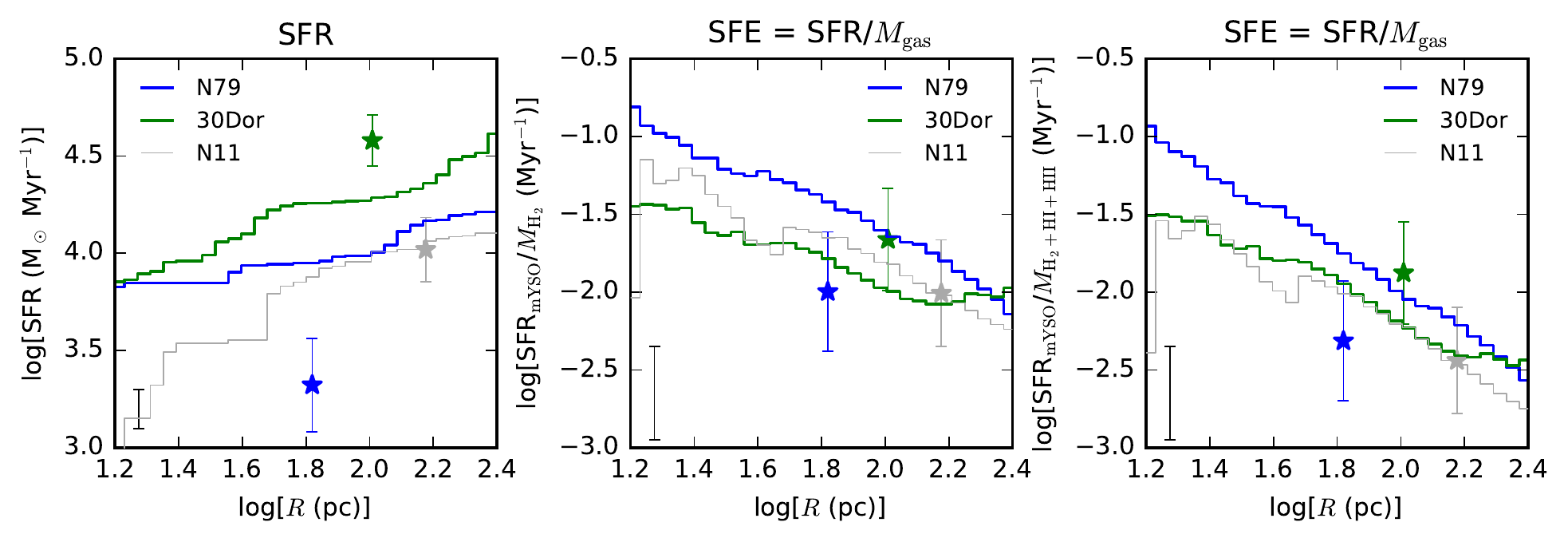} 
\caption{{\bf Star formation properties: N79 versus N11 and 30 Doradus.} {\bf (a)}: The SFR as measured by MYSO counting, SFR$_\m{MYSO}$, in apertures of radius $R$ centered on H72.97-69.39 (RA = 72.972, DEC = -69.391), N11 (RA = 74.227, DEC = -66.368), and R136 (RA = 84.633, DEC = -69.092). The asterisks marks the SFR of the regions as measured by H$\alpha$, SFR$_\m{H\alpha}$, where the size $R$ of the \HII\ regions is defined to enclose 90\% of the total flux from the central clusters\cite{lopez_2014}. Error bars are dominated by multiplicity (SFR$_\m{MYSO}$) or stochastic sampling of the IMF (SFR$_\m{H\alpha}$; see Methods). {\bf (b):} The inverse of the gas depletion time, SFR/$M_\m{gas}$, where the gas includes only the molecular component. The error bar shows the absolute uncertainty, dominated by systematic uncertainties in determining the molecular gas mass. However, the relative uncertainties are expected to be lower (see Methods). {\bf (c):}  Same as (b), but now including the molecular, neutral, and ionized components.}
\label{fig:gassfr}
\end{figure*}
  
We measure the star formation characteristics of N79 within an aperture of increasing size centered on H72.97-69.39 and compare this with 30 Doradus and N11. The total SFR$_\m{MYSO}$ is obtained by counting MYSOs and using an initial mass function (IMF) and characteristic age ($t_\star$\,$\lesssim$\,0.5 Myr; see Methods). We then compare SFR$_\m{MYSO}$ with the SFR measured through H$\alpha$ and 24\,$\mu$m emission, SFR$_\m{H\alpha}$ (see Methods), which allows us to compare the average SFR over the past $\sim$\,0.5 Myr and $\sim$\,4 Myr, respectively. Figure 4a shows that N79 matches the current SFR$_\m{MYSO}$ of N11, while being a factor of $\sim$\,2 lower than 30 Doradus. N11 is currently exhibiting a second burst of star formation\cite{walborn_1992} and shows SFR$_\m{MYSO}$\,$\approx$\,SFR$_\m{H\alpha}$ , thus sustaining its average SFR over the past $\sim$\,4 Myr. 30 Doradus shows SFR$_\m{MYSO}$\,$\textless$\,SFR$_\m{H\alpha}$, consistent with its inferred star formation history, which dramatically accelerated roughly $\sim$\,7 Myrs ago, peaked around 1\,-\,3 Myrs ago, and is currently decelerating\cite{cignoni_2015}. Conversely, the SFR in N79 has significantly {\em increased} over the past few Myr (SFR$_\m{MYSO}$\,$\textgreater$\,SFR$_\m{H\alpha}$ ) and has yet to reach its peak star formation activity. N79 may therefore be in a similar accelerating star formation phase 30 Doradus was $\sim$\,7 Myrs ago. \

By dividing the measured SFR with the total gas reservoir surrounding the star forming complexes, we obtain the inverse of the local gas depletion timescale, SFR/$M_\mathrm{gas}$. This quantity provides a measure of the timescale to exhaust the available gas reservoir at the current SFR (assuming all gas would be converted into stars). While the molecular clouds in the LMC are associated with \HI\ envelopes\cite{fukui_2009}, it is unclear which fraction of $M_\mathrm{\HI}$ will eventually be available for star formation. Therefore, we consider two cases. First, we take $M_\mathrm{gas}$ = $M_\mathrm{H_\mathrm{2}}$, i.e., we only take into account the molecular (dust-based) material (Fig. 4b). Second, we assume $M_\mathrm{gas}$ = $M_\mathrm{H_\mathrm{2}}$ + $M_\mathrm{\HI}$ + $M_\mathrm{\HII}$. By combining the molecular (dust-based), neutral, and ionized gas, we attempt to estimate an upper limit to the available gas reservoir for star formation, while tracing gas which may have already been disrupted/dissociated by the ionizing radiation of massive stars. In both cases, it becomes apparent that N79 is the most efficient site of {\em current} massive star formation, exceeding N11 and 30 Doradus by a factor of $\sim$\,2.0 - 2.5.

The total SFR in the LMC from MYSOs is 1.8\,$\times$\,10$^5$\,$M_\odot$\,Myr$^{-1}$ versus 2.6\,$\times$\,10$^5$\,$M_\odot$\,Myr$^{-1}$ measured through H$\alpha$ (see Methods). A percentage of 18\%, 9\%, and 7\% of the total SFR$_\m{MYSO}$ originates from a $\sim$\,0.25 kpc radius centered on 30 Doradus, N79, and N11, respectively (a 0.25 kpc radius area subtends only 1/400 of the total \HI\ disk\cite{staveley-smith_2003}). These numbers will likely increase for N79, and decrease for both 30 Doradus and N11 (see above). While the absolute SFR$_\m{MYSO}$ of N79 and N11 do not differ significantly at $R$\,$\gtrsim$\,50 pc (Fig. 4), the star formation efficiency of N79 (through MYSOs) and 30 Doradus (through H$\alpha$) are elevated compared to N11. This may suggest that the location of 30 Doradus and N79 on the leading edges of Arm E and Arm W positively influences the local star formation efficiency. 

Could the central object in N79, H72.97-69.39, eventually evolve into a SSC like R136? The total luminosity of R136, $L_\m{tot}$\,$\sim$\,7.0$\times$\,10$^7$ $L_\odot$\cite{malamuth_1994}, is currently at least an order of magnitude higher than H72.97-69.39, $L_\m{tot}$\,$\sim$\,2.2$\times$\,10$^6$ $L_\odot$ (Fig. 3b). With a formation period of 5\,-\,10\,Myr\cite{cignoni_2015}, an average SFR of $\sim$\,1\,-\,2\,$\times$\,10$^4$ M$_\odot$\,Myr$^{-1}$ is needed to create a 10$^5$ $M_\odot$ stellar cluster, which is a factor of $\sim$\,1.5\,-\,3.0 higher than currently observed at the heart of N79 (Figure 4). However for MYSOs, $t_\star$ may be lower than 0.5 Myr\cite{mottram_2011}, which would increase our SFR$_\m{MYSO}$ estimates through SFR$_\m{MYSO}$\,$\propto$\,$t_\star^{-1}$ (see Methods). The properties of the surrounding gas reservoir also play a role in developing H72.97-69.39. If we assume that the (molecular) gas in N79 is gravitationally collapsing together with a formation timescale of 5\,-\,10\,Myr, a star formation efficiency per free-fall time $\epsilon_\m{ff}$\,$\sim$\,0.27\,-\,0.75 (0.50 - 0.90 when limiting to the molecular gas) would need to be attained (see Methods). In this regard, values of $\epsilon_\m{ff}$\,$\textgreater$\,0.50 have been observed with recent large-scale surveys of individual GMCs in the LMC and Milky Way\cite{lee_2016,vutisalchavakul_2016,ochsendorf_2017,murray_2011},  while extraordinary high star formation efficiencies have been quoted for more distant SSCs\cite{turner_2015}. However, we note that observed values of $\epsilon_\m{ff}$ of individual GMCs extend over several orders of magnitude. Plus, stellar feedback may disrupt the cluster formation process, although the exact effects of feedback on massive protoclusters remain unclear\cite{ginsburg_2016}. Finally, the formation timescale of SSCs may be much smaller than our assumed 5\,-10\,Myr\cite{crowther_2016}. All of these effects may limit the final cluster mass. In this regard, detailed follow-up observations with the {\em Atacama Large Milimeter Array} (ALMA) and the upcoming {\em  James Webb Space Telescope} (JWST) are needed to establish if H72.97-69.39 could evolve into a SSC like R136 ($\sim$\,10$^5$ $M_\odot$), or a less-massive counterpart similar to the Arches and Quintuplet clusters near the Galactic center ($\sim$\,10$^4$ $M_\odot$\cite{figer_1999}).

The formation of very massive stars and SSCs is poorly understood\cite{krumholz_2014c}. In this regard, our findings on N79 and H72.97-69.39 highlights the importance of high-resolution IR observations to unveil the earliest phases of extreme star formation. The unique location of 30 Doradus and N79 suggest that the crossroads of spiral arms and galactic bars-ends may provide the right physical conditions to create massive clusters\cite{athanassoula_1992}. However, other factors that may play a role are the area-normalized SFR of a galaxy\cite{johnson_2017}, accretion flows\cite{turner_2015}, or tidal interactions\cite{bekki_2007}: observations suggest that R136 formed after a recent collision of distinct \HI\ flows, which were initially induced by the last LMC\,-\,SMC interaction $\sim$\,0.2 Gyr ago\cite{fukui_2017}. Because of the proximity and face-on orientation of the LMC, ALMA and JWST will allow to spatially resolve the formation of this candidate SSC down to $\lesssim$\,0.02 pc scales, which may reveal in exquisite detail how extreme star formation ignites and proceeds in the current and high-redshift Universe.

\subsection{Author correspondence.} 
Correspondence and request for materials should be directed to B. B. Ochsendorf. \\

\subsection{Author contributions.} B.\,B.\,O. performed the analysis, coordinated collaboration, and wrote the manuscript. O.\,N. helped characterizing H72.97-69.39. M.\,M. and O.\,C.\,J. helped with the creation of the MYSO catalog and estimates of source contamination. H.\,Z., J.\,B., R.\,I., and M.\,R. provided help with the interpretation of the results and implications. \\ 

\newpage
\begin{methods}

\subsection{LMC surveys.} In this work we have made use of various galaxy-wide surveys of the LMC:

\begin{enumerate}
\item {\em Atomic gas:} 21 cm data from the Australian Telescope Compact Array and Parkes 64 m radio Telescope map\cite{kim_2003}.
\item {\em Molecular gas:} $^{12}$CO\,(1-0) data from the Magellanic Mopra Assessment\cite{wong_2011} (MAGMA) Data Release 2 (resolution 45").
\item {\em Ionized gas:} H$\alpha$ from the Southern H-Alpha Sky Survey Atlas\cite{gaustad_2001} (SHASSA) was used for calculating the ionized gas mass. The H$\alpha
$ image displayed in Figure 2a stems from the Magellanic Clouds Emission Line Survey\cite{smith_1998} (MCELS), which has higher resolution compared to SHASSA but is not calibrated nor continuum subtracted.
\item {\em Infrared:} 3.6, 4.5, 5.8, and 8.0 $\mu$m mid-IR data from Spitzer's Surveying the Agents of a Galaxy's Evolution\cite{meixner_2006} (SAGE) and 70, 160, 250, 350, and 500 $\mu$m far-IR data from the Herschel Inventory of the Agents of Galaxy Evolution\cite{meixner_2013} (HERITAGE). 
\end{enumerate}\

\subsection{MYSO selection \& completeness.}
We have compiled a catalog of (highly) probable YSOs by combining the results of galaxy-wide searches of YSO candidates\cite{whitney_2008, gruendl_2009,seale_2014,seale_2009} using SAGE and HERITAGE data. The creation of the catalogue is explained in detail elsewhere\cite{ochsendorf_2016}, but the essential points are discussed here as well.

YSO candidates are identified through careful selection criteria (e.g., color-magnitude cuts, morphological inspection) tailored to minimize contamination from sources such as planetary nebulae, evolved stars, and background galaxies. Contamination estimates range from $\sim$\,55\%\cite{whitney_2008} to $\sim$\,10\%\cite{seale_2014}. This means that in regions of high source density (such as the stellar bar), a relatively large amount of false source candidates can be expected (see Figure 1b). However, contamination levels vary between the faint and bright end of the YSO distribution, as faint YSOs overlap more with the aforementioned contaminants in color-magnitude space compared to their luminous (i.e., higher-mass) counterparts. For high mass YSO candidates, the contamination from evolved stars and background galaxies is shown to be $\lesssim$20\%\cite{jones_2017}. However, in star-forming regions the contamination becomes negligible ($\textless$\,1\%) once SED fitting criteria has been applied. 

We combine the high-probable YSO candidates from the aforementioned studies\cite{whitney_2008, gruendl_2009,seale_2014} and subsequently fit their spectral energy distributions with YSO models\cite{robitaille_2006}.  These models (2\,$\times$\,10$^5$ in total) cover a wide range of physical parameters for different stages in the YSO evolutionary path, often divided in Stage 1 (least evolved), 2, and 3 (most evolved). The stringent color cuts used to separate out YSOs from fore- and background contaminations renders our census of Stage 2 and Stage 3 sources incomplete. However, these sources are largely irrelevant to this work since we aim to probe youngest population of YSOs, i.e., the earliest stages of (massive) star formation. 

The age of Stage 1 MYSOs is estimated at 0.5 Myr, which is the most recent value obtained for the observationally-derived `Class 1' low-mass sources (which largely overlap with the theoretically-based `Stage 1' sources\cite{heiderman_2015, heyer_2016}) in the Gould's Belt\cite{dunham_2015}. It is not clear whether this value applies to massive stars; the absolute durations of the starless and active star-forming phases for massive protostars is highly uncertain\cite{battersby_2017}. In addition, the accreting phase for massive protostars may decrease with luminosity, possibly reaching 0.1 Myr for a 10$^5$ $L_\odot$ star\cite{mottram_2011}. Indeed, massive stars are expected to evolve more quickly than their lower-mass counterparts, and the assumed age may therefore represent an upper limit to the age of these systems. A younger age would impact our results by {\em increasing} our SFR through SFR\,$\propto$\,$t_\star^{-1}$, where $t_\m{\star}$ is the age of the YSO population.

Completeness of the YSO catalogues has been evaluated through false source extraction tests for both the SAGE\cite{gruendl_2009} and HERITAGE\cite{meixner_2013} data, and conclude that our catalogue of YSOs should be complete for Stage 1 MYSOs of $M$\,$\textgreater$\,8 M$_\odot$ ($L$\,$\gtrsim$\,10$^3$ $L_\mathrm{\odot}$). We set the photometric errors to 10\% in the 2MASS, IRAC, and MIPS bands, which allows to account for multiple sources of error (systematic, calibration, variability, photon-counting)\cite{whitney_2008}. However, as PAH emission is not incorporated into the SED models (which will alter the emission in the 3.6, 5.8, and 8.0 $\mu$m bands compared to the models), we relax our constraints and adjust the error bars in these bands to 20\%, 30\%, and 40\%, respectively, corresponding to the intrinsic strengths of the PAH bands\cite{carlson_2011}. For the HERITAGE data, other sources of uncertainty were considered as well (background, PSF shapes)\cite{meixner_2013}: typical uncertainties reported in the HERITAGE catalog are of order 5\% - 20\%.  We only consider `well-fitted' sources, i.e. those yielding reduced chi square of $\chi_\mathrm{red}^2$\,$\leq$\,5 with the YSO models. By choosing to remove sources above a fixed reduced chi-squared value further decreases our source contamination. However, poorly-fitted sources may also arise because of a bad data point, a mismatch in photometric bands because of variability, multiplicity, or inadequacies of the YSO models\cite{whitney_2008}. Therefore, we stress that a poor fit does not necessarily mean that the object is not a YSO. We ultimately end with a catalogue of 693 Stage 1 MYSOs ($M$\,$\textgreater$\,8 M$_\odot$) across the LMC. \\

\subsection{Spectral energy distribution of H72.97-69.39.}
The spectral energy distribution of the central luminous source H72.97-69.39 was compiled from the InfraRed Survey Facility (IRSF)\cite{kato_2007}, WISE\cite{wright_2010}, SAGE\cite{meixner_2006}, and HERITAGE\cite{meixner_2013}. Its exceptional brightness and extended morphology causes the YSO model to severely underestimates its far-IR flux\cite{sewilo_2010}. Instead, we use a simple two-temperature modified blackbody (MBB) function, where the temperature $T$, spectral index $\beta$, and scaling parameters are left as free parameters\cite{gordon_2014}. The temperature of the hot dust component, peaking at $\sim$\,10 $\mu$m, radiates at $T_\mathrm{1}$ = 300 K (with $\beta_\mathrm{1}$ = 0.8), while the cold component peaking at $\sim$\,50 $\mu$m has $T_\mathrm{2}$ = 60 K (with $\beta_\mathrm{2}$ = 0.8). From this, we obtain a total infrared luminosity of $L_\m{IR}$ = 2.2\,$\times$\,10$^6$ $M_\m{sun}$. \\

\subsection{Mass determination.} The mass in N79, N11, and 30 Doradus in the various phases of the ISM (Fig. 4) is estimated in the following ways:

\begin{figure*}[!h]
\centering
\includegraphics[width=15cm]{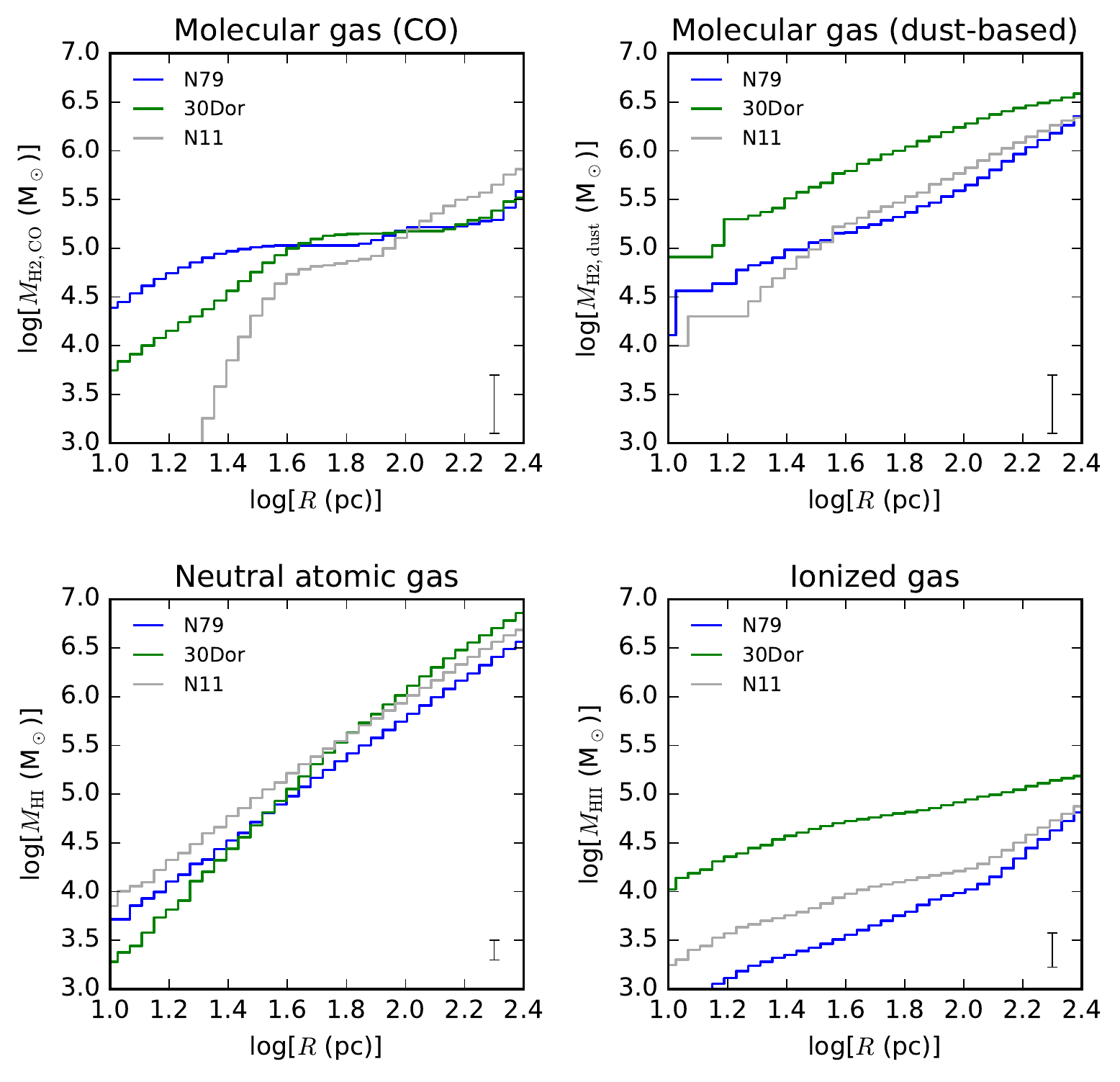} 
\caption{{\bf ISM properties of N79, N11, and 30 Doradus.} The molecular (CO-based and dust-based), neutral, and ionized gas in apertures of radius $R$ centered on N79 (RA = 72.972, DEC = -69.391), 30 Doradus (RA = 84.633, DEC = -69.092), and N11 (RA = 74.227, DEC = -66.368).}
\label{fig:gas}
\end{figure*}

\begin{enumerate}
\item {\em Neutral atomic mass}: assuming the \HI\ gas is optically thin, the column density is estimated through $N_\mathrm{HI}$ = $X_\mathrm{HI}$$W_\mathrm{HI}$, where $W_\mathrm{HI}$ is the integrated \HI\ intensity and $X_\mathrm{HI}$ = 1.82\,$\times$\,10$^{18}$ H cm$^2$/(K km s$^{-1}$) is the proportionality constant\cite{spitzer_1978}. This can subsequently be converted to gas surface density in $M_\odot$\,pc$^{-2}$ with $\Sigma_\mathrm{\HI}$ = 0.8\,$\times$10$^{-20}$ $N_\mathrm{\HI}$. We note that optically thick and/or cold \HI\ gas emits disproportionately compared to optically thin \HI. From absorption spectra it is known that the 21 cm line may be optically thick in the LMC\cite{marx-zimmer_2000}. Many studies have attempted to estimate the optical depth correction to the \HI\ mass, with differences of 10\% to 30\% reported compared to optically thin gas\cite{dickey_2000,lee_2015b}. Thus, we adopt an uncertainty of 0.1 dex for the \HI\ mass.
\item {\em CO-based molecular mass:} estimated through $M$ = $\alpha_\m{CO}$$L_\m{CO}$, where $L_\m{CO}$ is the CO luminosity and $\alpha_\m{CO}$ = 8.6 (K\,km\,s$^{-1}$\,pc$^2$)$^{-1}$ is the proportionality constant appropriate for the LMC\cite{bolatto_2013}. The $\alpha_\m{CO}$ factor is expected to be accurate within $\sim$\,0.3 dex\cite{bolatto_2013}.
\item {\em Dust-based molecular mass:} obtained by subtracting from the gas surface density, based on far-infrared dust emission (modeled with a single temperature blackbody modified by a broken power-law emissivity\cite{gordon_2014}), the surface density of atomic hydrogen. The creation of this map is discussed in detail elsewhere\cite{jameson_2016}, but the main caveats are reiterated here. First, the optical thin limit was used to convert \HI\ intensity to column density (see above). Second, it was assumed that the gas-to-dust ratio in the diffuse and atomic gas is the same as in molecular regions. However, there is mounting evidence that this gas-to-dust ratio changes between different phases of the ISM, with lower gas-to-dust ratios in the dense phase compared to the diffuse phase\cite{roman-duval_2014}. Both optically thick \HI\ and a decrease in the gas-to-dust ratio may lead to an overestimation of the dust-based molecular gas. These effects introduce a systematic uncertainty of $\sim$\,0.3\,dex in the dust-based molecular gas estimate\cite{jameson_2016}. However, given that we are focussing on (dense) star forming regions only, these uncertainties will propagate similarly for 30 Doradus, N11, and N79, and therefore the relative uncertainties are expected to be smaller.
\item {\em Ionized gas}: \HII\ column density in cm$^{-2}$ can be obtained by estimating electron densities in different brightness regimes\cite{paradis_2011}, which can then be converted to gas surface density in $M_\odot$\,pc$^{-2}$ through $\Sigma_\m{\HII}$ = 0.8\,$\times$10$^{-20}$ $N_\m{\HII}$. The systematic uncertainty is estimated at $\sim$\,0.2 dex. This mainly stems from the conversion from H$\alpha$ intensity to emission measure, which depends on the assumed electron temperature, which can vary\cite{shaver_1983} within a factor of $\sim$\,2, leading to a $\sim$\,50\% difference ($\sim$\,0.2 dex) in conversion from H$\alpha$ intensity to emission measure\cite{dickinson_2003}.

\end{enumerate}
   
Figure 5 shows the molecular (CO and dust-based), neutral, and ionized gas mass. Note that the total CO-based molecular mass in 30 Doradus and N79  are very similar within 100 pc. However, the dust-based molecular mass in 30 Doradus exceeds the CO-based material by almost an order of magnitude, indicating that the bulk of molecular material around 30 Doradus resides in the `CO-dark' phase, possibly through the local intense radiation field from R136. The ionized gas content around the clusters both reflect the mass (i.e., ionizing photon budget) of the central cluster and the evolutionary state (i.e., embeddedness) of the region.   \\

\subsection{Virial analysis.}
 
The virial parameter, $\alpha_\m{vir}$ = 5$\sigma_\m{v}^2$$R$/($GM$)\cite{bertoldi_1992}, where $\sigma_\m{v}^2$ is the luminosity-weighted (one-dimensional) CO velocity dispersion, $G$ is the gravitational constant, and $M$ the CO mass, can be used to determine whether a cloud (complex) is bound and can undergo collapse, or is unbound, and may expand and dissolve back into the ISM. The critical virial parameter is $\alpha_\m{cr}$\,$\simeq$\,2, with bound clouds having $\alpha_\m{cr}$\,$\leq$\,2. However, lower values for $\alpha_\m{cr}$ are possible in the case of strong magnetic fields\cite{kauffmann_2013}.
 
Figure 6 shows that the molecular gas in the entire N11 region contains extreme high $\alpha_\m{vir}$, which is to be expected given the evolved state of the region, with an expanding ring of material moving outward from the central `hole', where an earlier generation of massive stars appear to have been born\cite{rosado_1996} (note that individual cloud fragments may still be collapsing). 30 Doradus also contains elevated $\alpha_\m{vir}$; higher-resolution studies show that the CO gas in 30 Doradus has elevated CO linewidths, probably due to the highly energetic environmental conditions within 30 Doradus\cite{nayak_2016}. Conversely, N79 reveals sub-critical $\alpha_\m{vir}$ throughout the majority part of the N79 cloud complex. This indicates that the N79 cloud complex is bound, and may be in a collapsing state. \\

\begin{figure*}[]
\centering
\includegraphics[width=7.5cm]{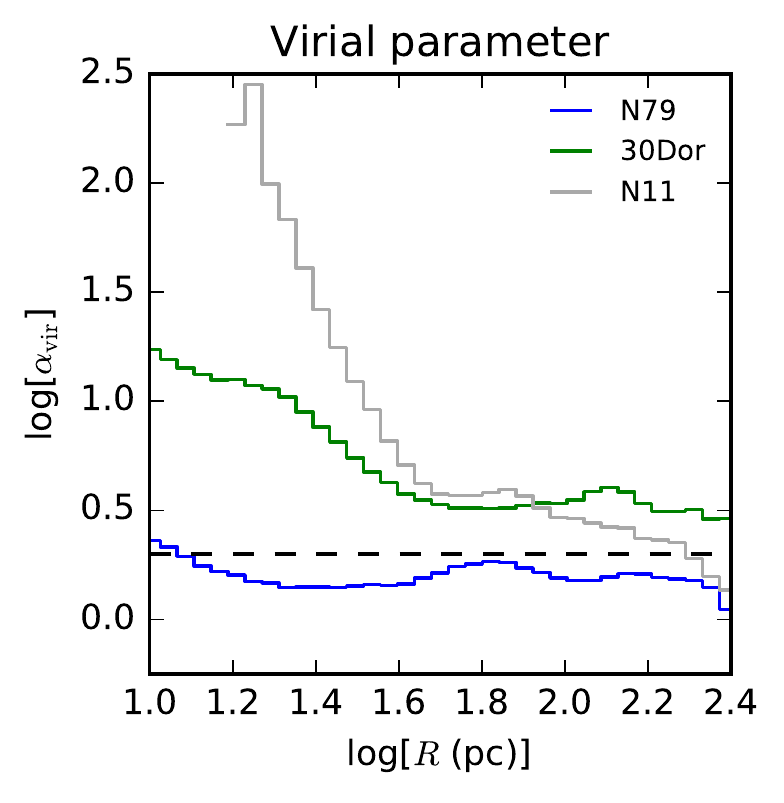} 
\caption{{\bf Virial analysis.} The virial parameter $\alpha_\m{vir}$ plotted versus radius $R$ in N79, N11, and 30 Doradus.}
\label{fig:virial}
\end{figure*}  

\subsection{Star formation rates (SFR) from MYSOs.} Young Stellar Objects (YSOs) can be used to obtain a direct measure of SFR through SFR = $N$(YSOs)\,$\times$\,$\langle{M}_\m{\star}\rangle$ / $t_\m{\star}$. Here, $\langle{M}_\m{\star}\rangle$ $\approx$ 0.5 is the average mass for a fully-sampled IMF\cite{kroupa_2001}. Whereas counting YSOs has previously been applied in studies of nearby molecular clouds, in the LMC we are limited to bright objects. Here, we assume that the luminosity of an MYSO is dominated by a single source. This assumption is motivated by observations of nearby star clusters (e.g., the Orion Trapezium\cite{ochsendorf_2016}). Subsequently, we use YSO models to estimate the mass of each individual source\cite{robitaille_2006} and multiply the source mass with an IMF\cite{kroupa_2001} to account for completeness. We choose 0.5 Myr for $t_\star$ (but see above). With these assumptions, the completeness limit ($M$\,$\textgreater$\,8 $M_\odot$) translates to a lower limit of SFR$_\m{MYSO}$\,$\sim$\,100 $M_\odot$ Myr$^{-1}$ which we can detect in our observations of the LMC.

Many of the sources in our MYSO sample will break into small clusters when observed at high resolution\cite{vaidya_2009,stephens_2017}. We estimate the uncertainty in our SFR$_\m{MYSO}$ measurement associated with multiplicity as follows. The reprocessed IR luminosity of the Orion Trapezium cluster would appear as a compact source at the resolution of our LMC IR maps. In the Orion Trapezium, the main ionizing source $\theta^1$ Ori C emits $\simeq$\,50\% of the total luminosity of the cluster\cite{vacca_1996,o'dell_1993,simon-diaz_2003}. The IR luminosity would thus overestimate the luminosity of a single most massive source by a factor of $\sim$\,0.3 dex, which translates to an error of $\sim$\,0.1 dex in mass\cite{mottram_2011}. Therefore, we adopt 0.1 dex as our systematic uncertainty in SFR$_\m{MYSO}$; note that this assumes that the evolutionary tracks used in the YSO models are correct\cite{robitaille_2008}.\\

\subsection{Star formation rates (SFR) from H$\alpha$\,+\,24 $\mu$m.} We convolve the 24 $\mu$m map (from SAGE\cite{meixner_2013}) to the resolution of our H$\alpha$ map (resolution 0.8'; from SHASSA\cite{gaustadt_2001}), and correct the H$\alpha$ emission for extinction using the 24 $\mu$m emission. We then transform the combined H$\alpha$ and 24 $\mu$m luminosity, $L$\,(H$\alpha$) and $L$\,(24$\mu$m), to a SFR\cite{calzetti_2007} :

\begin{equation}	
\begin{split}
\m{SFR}_\m{H\alpha}\,(M_\m{\odot}\,\m{yr}^{-1}) = 5.3\times10^{-42} [L\,(\m{H}\alpha) + \,0.031\,L\,(24\,\mu\m{m})].
\end{split}
\label{eq:sfr}
\end{equation}

We note that Eq. \ref{eq:sfr} assumes a fully-sampled IMF, which can be attained by averaging over large spatial scales\cite{krumholz_2014} such that each phase of star formation is probed. When studying star formation on smaller scales\cite{kruijssen_2014} these assumption may break down, which introduces stochastic effects that mainly affect the high-end part of the IMF. To account for this, we use the tool `Stochastically Lighting Up Galaxies' (SLUG\cite{krumholz_2014}) to estimate to which extend stochastic sampling of the IMF affect our measured SFR. This process is fully described elsewhere\cite{ochsendorf_2017}. The total SFR in the LMC measured through H$\alpha$ equals\cite{kennicutt_1995} 2.6\,$\times$\,10$^5$\,$M_\odot$\,Myr$^{-1}$. \\ 

\subsection{Star formation efficiency per free-fall time.}
Can H72.97-69.39 become a 10$^5$ M$_\odot$ star cluster? The current state of the (molecular) gas in N79 reveals that it is gravitationally bound and prone to further collapse (Figure 6). We thus assume that the gas collapses on its gravitational timescale, the free-fall time $\tau_\m{ff}$ = $\sqrt{3\pi/32G\rho}$, where $\rho$ = $M_\m{cloud}$/(4/3$\pi$$R_\m{cloud}^3$) is the mean density, $M_\m{cloud}$ the cloud mass, $R_\m{cloud}$ the cloud radius, and $G$ the gravitational constant. We adopt a formation timescale of 5 - 10 Myr, motivated by the star formation history of the NGC2070 region\cite{cignoni_2015} (including R136). In this timescale, we calculate that a total of 1.3\,-\,3.7\,$\times$10$^5$ of mass (molecular and atomic) can collapse to the centre of the N79 cloud from $R_\m{cloud}$\,$\lesssim$\,30\,-\,60 pc. Thus, to create a 10$^5$ M$_\odot$ stellar cluster, a star formation efficiency per free-fall time (i.e., the fraction of mass that is transformed into stars\cite{krumholz_2005} during $\tau_\m{ff}$) of $\epsilon_\m{ff}$\,$\sim$\,0.27\,-0.75 (0.50 - 0.90 when only considering molecular gas) would be required. 

The question whether or not a 10$^5$ M$_\odot$ cluster is massive enough to eventually become a globular cluster ($\gtrsim$\,10$^5$\,M$_\odot$)\cite{portegies-zwart_2010} is not clear, given that young stellar clusters may lose a significant amount of mass through supernovae, stellar winds, and stripping\cite{d'ercole_2008,bekki_2011,schaerer_2011}. \\

\subsection{Data availability statement.}
The data and analysis scripts that support the plots within this paper and other findings of this study will be made available through a public repository: \\ www.github.com/bramochsendorf. Additional requests can be directed to the corresponding author.


\vspace{1.5cm}


\begin{thebibliography}{10}

\expandafter\ifx\csname url\endcsname\relax
  \def\url#1{\texttt{#1}}\fi
\expandafter\ifx\csname urlprefix\endcsname\relax\def\urlprefix{URL }\fi
\providecommand{\bibinfo}[2]{#2}
\providecommand{\eprint}[2][]{\url{#2}}

\bibitem{barger_1998}
\bibinfo{author}{{Barger}, A.~J.} \emph{et~al.}
\newblock \bibinfo{title}{{Submillimetre-wavelength detection of dusty
  star-forming galaxies at high redshift}}.
\newblock \emph{\bibinfo{journal}{\nat}} \textbf{\bibinfo{volume}{394}},
  \bibinfo{pages}{248--251} (\bibinfo{year}{1998}).
\newblock \eprint{astro-ph/9806317}.

\bibitem{turner_2009}
\bibinfo{author}{{Turner}, J.~L.}
\newblock \bibinfo{title}{{Extreme Star Formation}}.
\newblock \emph{\bibinfo{journal}{Astrophysics and Space Science Proceedings}}
  \textbf{\bibinfo{volume}{10}}, \bibinfo{pages}{215} (\bibinfo{year}{2009}).
\newblock \eprint{1009.1416}.

\bibitem{kruijssen_2015}
\bibinfo{author}{{Kruijssen}, J.~M.~D.}
\newblock \bibinfo{title}{{Globular clusters as the relics of regular star
  formation in `normal' high-redshift galaxies}}.
\newblock \emph{\bibinfo{journal}{\mnras}} \textbf{\bibinfo{volume}{454}},
  \bibinfo{pages}{1658--1686} (\bibinfo{year}{2015}).
\newblock \eprint{1509.02163}.

\bibitem{walborn_1991}
\bibinfo{author}{{Walborn}, N.~R.}
\newblock \bibinfo{title}{{The Starburst Region 30 Doradus}}.
\newblock In \bibinfo{editor}{{Haynes}, R.} \& \bibinfo{editor}{{Milne}, D.}
  (eds.) \emph{\bibinfo{booktitle}{The Magellanic Clouds}}, vol.
  \bibinfo{volume}{148} of \emph{\bibinfo{series}{IAU Symposium}},
  \bibinfo{pages}{145} (\bibinfo{year}{1991}).

\bibitem{de_vaucouleurs_1972}
\bibinfo{author}{{de Vaucouleurs}, G.} \& \bibinfo{author}{{Freeman}, K.~C.}
\newblock \bibinfo{title}{{Structure and dynamics of barred spiral galaxies, in
  particular of the Magellanic type}}.
\newblock \emph{\bibinfo{journal}{Vistas in Astronomy}}
  \textbf{\bibinfo{volume}{14}}, \bibinfo{pages}{163--294}
  (\bibinfo{year}{1972}).

\bibitem{staveley-smith_2003}
\bibinfo{author}{{Staveley-Smith}, L.}, \bibinfo{author}{{Kim}, S.},
  \bibinfo{author}{{Calabretta}, M.~R.}, \bibinfo{author}{{Haynes}, R.~F.} \&
  \bibinfo{author}{{Kesteven}, M.~J.}
\newblock \bibinfo{title}{{A new look at the large-scale HI structure of the
  Large Magellanic Cloud}}.
\newblock \emph{\bibinfo{journal}{\mnras}} \textbf{\bibinfo{volume}{339}},
  \bibinfo{pages}{87--104} (\bibinfo{year}{2003}).
\newblock \eprint{astro-ph/0210501}.

\bibitem{bekki_2007}
\bibinfo{author}{{Bekki}, K.} \& \bibinfo{author}{{Chiba}, M.}
\newblock \bibinfo{title}{{Dynamical Influences of the Last Magellanic
  Interaction on the Magellanic Clouds}}.
\newblock \emph{\bibinfo{journal}{\pasa}} \textbf{\bibinfo{volume}{24}},
  \bibinfo{pages}{21--29} (\bibinfo{year}{2007}).
\newblock \eprint{astro-ph/0603812}.

\bibitem{de_boer_1998}
\bibinfo{author}{{de Boer}, K.~S.}, \bibinfo{author}{{Braun}, J.~M.},
  \bibinfo{author}{{Vallenari}, A.} \& \bibinfo{author}{{Mebold}, U.}
\newblock \bibinfo{title}{{Bow-shock induced star formation in the LMC?}}
\newblock \emph{\bibinfo{journal}{\aap}} \textbf{\bibinfo{volume}{329}},
  \bibinfo{pages}{L49--L52} (\bibinfo{year}{1998}).
\newblock \eprint{astro-ph/9711052}.

\bibitem{murray_2011}
\bibinfo{author}{{Murray}, N.}
\newblock \bibinfo{title}{{Star Formation Efficiencies and Lifetimes of Giant
  Molecular Clouds in the Milky Way}}.
\newblock \emph{\bibinfo{journal}{\apj}} \textbf{\bibinfo{volume}{729}},
  \bibinfo{pages}{133} (\bibinfo{year}{2011}).
\newblock \eprint{1007.3270}.

\bibitem{ochsendorf_2016}
\bibinfo{author}{{Ochsendorf}, B.~B.}, \bibinfo{author}{{Meixner}, M.},
  \bibinfo{author}{{Chastenet}, J.}, \bibinfo{author}{{Tielens}, A.~G.~G.~M.}
  \& \bibinfo{author}{{Roman-Duval}, J.}
\newblock \bibinfo{title}{{The Location, Clustering, and Propagation of Massive
  Star Formation in Giant Molecular Clouds}}.
\newblock \emph{\bibinfo{journal}{\apj}} \textbf{\bibinfo{volume}{832}},
  \bibinfo{pages}{43} (\bibinfo{year}{2016}).
\newblock \eprint{1609.03522}.

\bibitem{henize_1956}
\bibinfo{author}{{Henize}, K.~G.}
\newblock \bibinfo{title}{{Catalogues of H{$\alpha$}-emission Stars and Nebulae
  in the Magellanic Clouds.}}
\newblock \emph{\bibinfo{journal}{\apjs}} \textbf{\bibinfo{volume}{2}},
  \bibinfo{pages}{315} (\bibinfo{year}{1956}).

\bibitem{madden_1997}
\bibinfo{author}{{Madden}, S.~C.}, \bibinfo{author}{{Poglitsch}, A.},
  \bibinfo{author}{{Geis}, N.}, \bibinfo{author}{{Stacey}, G.~J.} \&
  \bibinfo{author}{{Townes}, C.~H.}
\newblock \bibinfo{title}{{[C II] 158 Micron Observations of IC 10: Evidence
  for Hidden Molecular Hydrogen in Irregular Galaxies}}.
\newblock \emph{\bibinfo{journal}{\apj}} \textbf{\bibinfo{volume}{483}},
  \bibinfo{pages}{200--209} (\bibinfo{year}{1997}).

\bibitem{jameson_2016}
\bibinfo{author}{{Jameson}, K.~E.} \emph{et~al.}
\newblock \bibinfo{title}{{The Relationship Between Molecular Gas, H I, and
  Star Formation in the Low-mass, Low-metallicity Magellanic Clouds}}.
\newblock \emph{\bibinfo{journal}{\apj}} \textbf{\bibinfo{volume}{825}},
  \bibinfo{pages}{12} (\bibinfo{year}{2016}).
\newblock \eprint{1510.08084}.

\bibitem{seale_2014}
\bibinfo{author}{{Seale}, J.~P.} \emph{et~al.}
\newblock \bibinfo{title}{{Herschel Key Program Heritage: a Far-Infrared Source
  Catalog for the Magellanic Clouds}}.
\newblock \emph{\bibinfo{journal}{\aj}} \textbf{\bibinfo{volume}{148}},
  \bibinfo{pages}{124} (\bibinfo{year}{2014}).

\bibitem{mottram_2011}
\bibinfo{author}{{Mottram}, J.~C.} \emph{et~al.}
\newblock \bibinfo{title}{{The RMS Survey: The Luminosity Functions and
  Timescales of Massive Young Stellar Objects and Compact H II Regions}}.
\newblock \emph{\bibinfo{journal}{\apjl}} \textbf{\bibinfo{volume}{730}},
  \bibinfo{pages}{L33} (\bibinfo{year}{2011}).
\newblock \eprint{1102.4702}.

\bibitem{zinnecker_2007}
\bibinfo{author}{{Zinnecker}, H.} \& \bibinfo{author}{{Yorke}, H.~W.}
\newblock \bibinfo{title}{{Toward Understanding Massive Star Formation}}.
\newblock \emph{\bibinfo{journal}{\araa}} \textbf{\bibinfo{volume}{45}},
  \bibinfo{pages}{481--563} (\bibinfo{year}{2007}).
\newblock \eprint{0707.1279}.

\bibitem{walborn_1992}
\bibinfo{author}{{Walborn}, N.~R.} \& \bibinfo{author}{{Parker}, J.~W.}
\newblock \bibinfo{title}{{Two-stage starbursts in the Large Magellanic Cloud -
  N11 as a once and future 30 Doradus}}.
\newblock \emph{\bibinfo{journal}{\apjl}} \textbf{\bibinfo{volume}{399}},
  \bibinfo{pages}{L87--L89} (\bibinfo{year}{1992}).

\bibitem{cignoni_2015}
\bibinfo{author}{{Cignoni}, M.} \emph{et~al.}
\newblock \bibinfo{title}{{Hubble Tarantula Treasury Project. II. The
  Star-formation History of the Starburst Region NGC 2070 in 30 Doradus}}.
\newblock \emph{\bibinfo{journal}{\apj}} \textbf{\bibinfo{volume}{811}},
  \bibinfo{pages}{76} (\bibinfo{year}{2015}).
\newblock \eprint{1505.04799}.

\bibitem{fukui_2009}
\bibinfo{author}{{Fukui}, Y.} \emph{et~al.}
\newblock \bibinfo{title}{{Molecular and Atomic Gas in the Large Magellanic
  Cloud. II. Three-dimensional Correlation Between CO and H I}}.
\newblock \emph{\bibinfo{journal}{\apj}} \textbf{\bibinfo{volume}{705}},
  \bibinfo{pages}{144--155} (\bibinfo{year}{2009}).
\newblock \eprint{0909.0382}.

\bibitem{malamuth_1994}
\bibinfo{author}{{Malumuth}, E.~M.} \& \bibinfo{author}{{Heap}, S.~R.}
\newblock \bibinfo{title}{{UBV stellar photometry of the 30 Doradus region of
  the large Magellanic Cloud with the Hubble Space Telescope}}.
\newblock \emph{\bibinfo{journal}{\aj}} \textbf{\bibinfo{volume}{107}},
  \bibinfo{pages}{1054--1066} (\bibinfo{year}{1994}).

\bibitem{lee_2016}
\bibinfo{author}{{Lee}, E.~J.}, \bibinfo{author}{{Miville-Desch{\^e}nes},
  M.-A.} \& \bibinfo{author}{{Murray}, N.~W.}
\newblock \bibinfo{title}{{Observational Evidence of Dynamic Star Formation
  Rate in Milky Way Giant Molecular Clouds}}.
\newblock \emph{\bibinfo{journal}{\apj}} \textbf{\bibinfo{volume}{833}},
  \bibinfo{pages}{229} (\bibinfo{year}{2016}).
\newblock \eprint{1608.05415}.

\bibitem{vutisalchavakul_2016}
\bibinfo{author}{{Vutisalchavakul}, N.}, \bibinfo{author}{{Evans}, N.~J., II}
  \& \bibinfo{author}{{Heyer}, M.}
\newblock \bibinfo{title}{{Star Formation Relations in the Milky Way}}.
\newblock \emph{\bibinfo{journal}{\apj}} \textbf{\bibinfo{volume}{831}},
  \bibinfo{pages}{73} (\bibinfo{year}{2016}).
\newblock \eprint{1607.06518}.

\bibitem{turner_2015}
\bibinfo{author}{{Turner}, J.~L.} \emph{et~al.}
\newblock \bibinfo{title}{{Highly efficient star formation in NGC 5253 possibly
  from stream-fed accretion}}.
\newblock \emph{\bibinfo{journal}{\nat}} \textbf{\bibinfo{volume}{519}},
  \bibinfo{pages}{331--333} (\bibinfo{year}{2015}).
\newblock \eprint{1503.05254}.

\bibitem{ginsburg_2016}
\bibinfo{author}{{Ginsburg}, A.} \emph{et~al.}
\newblock \bibinfo{title}{{Toward gas exhaustion in the W51 high-mass
  protoclusters}}.
\newblock \emph{\bibinfo{journal}{\aap}} \textbf{\bibinfo{volume}{595}},
  \bibinfo{pages}{A27} (\bibinfo{year}{2016}).
\newblock \eprint{1605.09402}.

\bibitem{crowther_2016}
\bibinfo{author}{{Crowther}, P.~A.} \emph{et~al.}
\newblock \bibinfo{title}{{The R136 star cluster dissected with Hubble Space
  Telescope/STIS. I. Far-ultraviolet spectroscopic census and the origin of He
  II {$\lambda$}1640 in young star clusters}}.
\newblock \emph{\bibinfo{journal}{\mnras}} \textbf{\bibinfo{volume}{458}},
  \bibinfo{pages}{624--659} (\bibinfo{year}{2016}).
\newblock \eprint{1603.04994}.

\bibitem{figer_1999}
\bibinfo{author}{{Figer}, D.~F.} \emph{et~al.}
\newblock \bibinfo{title}{{Hubble Space Telescope/NICMOS Observations of
  Massive Stellar Clusters near the Galactic Center}}.
\newblock \emph{\bibinfo{journal}{\apj}} \textbf{\bibinfo{volume}{525}},
  \bibinfo{pages}{750--758} (\bibinfo{year}{1999}).
\newblock \eprint{astro-ph/9906299}.

\bibitem{krumholz_2014c}
\bibinfo{author}{{Krumholz}, M.~R.}
\newblock \bibinfo{title}{{The Formation of Very Massive Stars}}.
\newblock In \bibinfo{editor}{{Vink}, J.~S.} (ed.)
  \emph{\bibinfo{booktitle}{Very Massive Stars in the Local Universe}}, vol.
  \bibinfo{volume}{412} of \emph{\bibinfo{series}{Astrophysics and Space
  Science Library}}, \bibinfo{pages}{43} (\bibinfo{year}{2015}).
\newblock \eprint{1403.3417}.

\bibitem{athanassoula_1992}
\bibinfo{author}{{Athanassoula}, E.}
\newblock \bibinfo{title}{{The existence and shapes of dust lanes in galactic
  bars}}.
\newblock \emph{\bibinfo{journal}{\mnras}} \textbf{\bibinfo{volume}{259}},
  \bibinfo{pages}{345--364} (\bibinfo{year}{1992}).

\bibitem{johnson_2017}
\bibinfo{author}{{Johnson}, L.~C.} \emph{et~al.}
\newblock \bibinfo{title}{{Panchromatic Hubble Andromeda Treasury. XVIII. The
  High-mass Truncation of the Star Cluster Mass Function}}.
\newblock \emph{\bibinfo{journal}{\apj}} \textbf{\bibinfo{volume}{839}},
  \bibinfo{pages}{78} (\bibinfo{year}{2017}).
\newblock \eprint{1703.10312}.

\bibitem{fukui_2017}
\bibinfo{author}{{Fukui}, Y.} \emph{et~al.}
\newblock \bibinfo{title}{{Formation of the young massive cluster R136
  triggered by tidally-driven colliding H i flows}}.
\newblock \emph{\bibinfo{journal}{\pasj}} \textbf{\bibinfo{volume}{69}},
  \bibinfo{pages}{L5} (\bibinfo{year}{2017}).
\newblock \eprint{1703.01075}.

\bibitem{van_der_marel_2014}
\bibinfo{author}{{van der Marel}, R.~P.} \& \bibinfo{author}{{Kallivayalil},
  N.}
\newblock \bibinfo{title}{{Third-epoch Magellanic Cloud Proper Motions. II. The
  Large Magellanic Cloud Rotation Field in Three Dimensions}}.
\newblock \emph{\bibinfo{journal}{\apj}} \textbf{\bibinfo{volume}{781}},
  \bibinfo{pages}{121} (\bibinfo{year}{2014}).
\newblock \eprint{1305.4641}.

\bibitem{ochsendorf_2017}
\bibinfo{author}{{Ochsendorf}, B.~B.}, \bibinfo{author}{{Meixner}, M.},
  \bibinfo{author}{{Roman-Duval}, J.}, \bibinfo{author}{{Rahman}, M.} \&
  \bibinfo{author}{{Evans}, N.~J., II}.
\newblock \bibinfo{title}{{What Sets the Massive Star Formation Rates and
  Efficiencies of Giant Molecular Clouds?}}
\newblock \emph{\bibinfo{journal}{\apj}} \textbf{\bibinfo{volume}{841}},
  \bibinfo{pages}{109} (\bibinfo{year}{2017}).
\newblock \eprint{1704.06965}.

\bibitem{lopez_2014}
\bibinfo{author}{{Lopez}, L.~A.} \emph{et~al.}
\newblock \bibinfo{title}{{The Role of Stellar Feedback in the Dynamics of H II
  Regions}}.
\newblock \emph{\bibinfo{journal}{\apj}} \textbf{\bibinfo{volume}{795}},
  \bibinfo{pages}{121} (\bibinfo{year}{2014}).
\newblock \eprint{1309.5421}.

\end{thebibliography}

\vspace{1.5cm}


\begin{thebibliography}{10}
\makeatletter
\addtocounter{\@listctr}{33}
\makeatother
\expandafter\ifx\csname url\endcsname\relax
  \def\url#1{\texttt{#1}}\fi
\expandafter\ifx\csname urlprefix\endcsname\relax\def\urlprefix{URL }\fi
\providecommand{\bibinfo}[2]{#2}
\providecommand{\eprint}[2][]{\url{#2}}


\bibitem{kim_2003}
\bibinfo{author}{{Kim}, S.} \emph{et~al.}
\newblock \bibinfo{title}{{A Neutral Hydrogen Survey of the Large Magellanic
  Cloud: Aperture Synthesis and Multibeam Data Combined}}.
\newblock \emph{\bibinfo{journal}{\apjs}} \textbf{\bibinfo{volume}{148}},
  \bibinfo{pages}{473--486} (\bibinfo{year}{2003}).

\bibitem{wong_2011}
\bibinfo{author}{{Wong}, T.} \emph{et~al.}
\newblock \bibinfo{title}{{The Magellanic Mopra Assessment (MAGMA). I. The
  Molecular Cloud Population of the Large Magellanic Cloud}}.
\newblock \emph{\bibinfo{journal}{\apjs}} \textbf{\bibinfo{volume}{197}},
  \bibinfo{pages}{16} (\bibinfo{year}{2011}).
\newblock \eprint{1108.5715}.

\bibitem{gaustad_2001}
\bibinfo{author}{{Gaustad}, J.~E.}, \bibinfo{author}{{McCullough}, P.~R.},
  \bibinfo{author}{{Rosing}, W.} \& \bibinfo{author}{{Van Buren}, D.}
\newblock \bibinfo{title}{{A Robotic Wide-Angle H{$\alpha$} Survey of the
  Southern Sky}}.
\newblock \emph{\bibinfo{journal}{\pasp}} \textbf{\bibinfo{volume}{113}},
  \bibinfo{pages}{1326--1348} (\bibinfo{year}{2001}).
\newblock \eprint{astro-ph/0108518}.

\bibitem{smith_1998}
\bibinfo{author}{{Smith}, R.~C.} \& \bibinfo{author}{{MCELS Team}}.
\newblock \bibinfo{title}{{The UM/CTIO Magellanic Cloud emission-line survey}}.
\newblock \emph{\bibinfo{journal}{\pasa}} \textbf{\bibinfo{volume}{15}},
  \bibinfo{pages}{163--64} (\bibinfo{year}{1998}).

\bibitem{meixner_2006}
\bibinfo{author}{{Meixner}, M.} \emph{et~al.}
\newblock \bibinfo{title}{{Spitzer Survey of the Large Magellanic Cloud:
  Surveying the Agents of a Galaxy's Evolution (SAGE). I. Overview and Initial
  Results}}.
\newblock \emph{\bibinfo{journal}{\aj}} \textbf{\bibinfo{volume}{132}},
  \bibinfo{pages}{2268--2288} (\bibinfo{year}{2006}).
\newblock \eprint{astro-ph/0606356}.

\bibitem{meixner_2013}
\bibinfo{author}{{Meixner}, M.} \emph{et~al.}
\newblock \bibinfo{title}{{The HERSCHEL Inventory of The Agents of Galaxy
  Evolution in the Magellanic Clouds, a Herschel Open Time Key Program}}.
\newblock \emph{\bibinfo{journal}{\aj}} \textbf{\bibinfo{volume}{146}},
  \bibinfo{pages}{62} (\bibinfo{year}{2013}).

\bibitem{whitney_2008}
\bibinfo{author}{{Whitney}, B.~A.} \emph{et~al.}
\newblock \bibinfo{title}{{Spitzer Sage Survey of the Large Magellanic Cloud.
  III. Star Formation and \~{}1000 New Candidate Young Stellar Objects}}.
\newblock \emph{\bibinfo{journal}{\aj}} \textbf{\bibinfo{volume}{136}},
  \bibinfo{pages}{18--43} (\bibinfo{year}{2008}).

\bibitem{gruendl_2009}
\bibinfo{author}{{Gruendl}, R.~A.} \& \bibinfo{author}{{Chu}, Y.-H.}
\newblock \bibinfo{title}{{High- and Intermediate-Mass Young Stellar Objects in
  the Large Magellanic Cloud}}.
\newblock \emph{\bibinfo{journal}{\apjs}} \textbf{\bibinfo{volume}{184}},
  \bibinfo{pages}{172--197} (\bibinfo{year}{2009}).
\newblock \eprint{0908.0347}.

\bibitem{seale_2009}
\bibinfo{author}{{Seale}, J.~P.} \emph{et~al.}
\newblock \bibinfo{title}{{The Evolution Of Massive Young Stellar Objects in
  the Large Magellanic Cloud. I. Identification and Spectral Classification}}.
\newblock \emph{\bibinfo{journal}{\apj}} \textbf{\bibinfo{volume}{699}},
  \bibinfo{pages}{150--167} (\bibinfo{year}{2009}).
\newblock \eprint{0904.1825}.

\bibitem{jones_2017}
\bibinfo{author}{{Jones}, O.~C.} \emph{et~al.}
\newblock \bibinfo{title}{{The SAGE-Spec Spitzer Legacy program: The life-cycle
  of dust and gas in the Large Magellanic Cloud. Point source classification
  III}}.
\newblock \emph{\bibinfo{journal}{ArXiv e-prints}}  (\bibinfo{year}{2017}).
\newblock \eprint{1705.02709}.

\bibitem{robitaille_2006}
\bibinfo{author}{{Robitaille}, T.~P.}, \bibinfo{author}{{Whitney}, B.~A.},
  \bibinfo{author}{{Indebetouw}, R.}, \bibinfo{author}{{Wood}, K.} \&
  \bibinfo{author}{{Denzmore}, P.}
\newblock \bibinfo{title}{{Interpreting Spectral Energy Distributions from
  Young Stellar Objects. I. A Grid of 200,000 YSO Model SEDs}}.
\newblock \emph{\bibinfo{journal}{\apjs}} \textbf{\bibinfo{volume}{167}},
  \bibinfo{pages}{256--285} (\bibinfo{year}{2006}).
\newblock \eprint{astro-ph/0608234}.

\bibitem{heiderman_2015}
\bibinfo{author}{{Heiderman}, A.} \& \bibinfo{author}{{Evans}, N.~J., II}.
\newblock \bibinfo{title}{{The Gould Belt 'MISFITS' Survey: The Real Solar
  Neighborhood Protostars}}.
\newblock \emph{\bibinfo{journal}{\apj}} \textbf{\bibinfo{volume}{806}},
  \bibinfo{pages}{231} (\bibinfo{year}{2015}).
\newblock \eprint{1503.06810}.

\bibitem{heyer_2016}
\bibinfo{author}{{Heyer}, M.} \emph{et~al.}
\newblock \bibinfo{title}{{The rate and latency of star formation in dense,
  massive clumps in the Milky Way}}.
\newblock \emph{\bibinfo{journal}{\aap}} \textbf{\bibinfo{volume}{588}},
  \bibinfo{pages}{A29} (\bibinfo{year}{2016}).
\newblock \eprint{1601.06875}.

\bibitem{dunham_2015}
\bibinfo{author}{{Dunham}, M.~M.} \emph{et~al.}
\newblock \bibinfo{title}{{Young Stellar Objects in the Gould Belt}}.
\newblock \emph{\bibinfo{journal}{\apjs}} \textbf{\bibinfo{volume}{220}},
  \bibinfo{pages}{11} (\bibinfo{year}{2015}).
\newblock \eprint{1508.03199}.

\bibitem{battersby_2017}
\bibinfo{author}{{Battersby}, C.}, \bibinfo{author}{{Bally}, J.} \&
  \bibinfo{author}{{Svoboda}, B.}
\newblock \bibinfo{title}{{The Lifetimes of Phases in High-mass Star-forming
  Regions}}.
\newblock \emph{\bibinfo{journal}{\apj}} \textbf{\bibinfo{volume}{835}},
  \bibinfo{pages}{263} (\bibinfo{year}{2017}).
\newblock \eprint{1702.02199}.

\bibitem{carlson_2011}
\bibinfo{author}{{Carlson}, L.~R.} \emph{et~al.}
\newblock \bibinfo{title}{{A Panchromatic View of NGC 602: Time-resolved Star
  Formation with the Hubble and Spitzer Space Telescopes}}.
\newblock \emph{\bibinfo{journal}{\apj}} \textbf{\bibinfo{volume}{730}},
  \bibinfo{pages}{78} (\bibinfo{year}{2011}).
\newblock \eprint{1012.3406}.

\bibitem{kato_2007}
\bibinfo{author}{{Kato}, D.} \emph{et~al.}
\newblock \bibinfo{title}{{The IRSF Magellanic Clouds Point Source Catalog}}.
\newblock \emph{\bibinfo{journal}{\pasj}} \textbf{\bibinfo{volume}{59}},
  \bibinfo{pages}{615--641} (\bibinfo{year}{2007}).

\bibitem{wright_2010}
\bibinfo{author}{{Wright}, E.~L.} \emph{et~al.}
\newblock \bibinfo{title}{{The Wide-field Infrared Survey Explorer (WISE):
  Mission Description and Initial On-orbit Performance}}.
\newblock \emph{\bibinfo{journal}{\aj}} \textbf{\bibinfo{volume}{140}},
  \bibinfo{pages}{1868--1881} (\bibinfo{year}{2010}).
\newblock \eprint{1008.0031}.

\bibitem{sewilo_2010}
\bibinfo{author}{{Sewi{\l}o}, M.} \emph{et~al.}
\newblock \bibinfo{title}{{The youngest massive protostars in the Large
  Magellanic Cloud}}.
\newblock \emph{\bibinfo{journal}{\aap}} \textbf{\bibinfo{volume}{518}},
  \bibinfo{pages}{L73} (\bibinfo{year}{2010}).
\newblock \eprint{1005.2592}.

\bibitem{gordon_2014}
\bibinfo{author}{{Gordon}, K.~D.} \emph{et~al.}
\newblock \bibinfo{title}{{Dust and Gas in the Magellanic Clouds from the
  HERITAGE Herschel Key Project. I. Dust Properties and Insights into the
  Origin of the Submillimeter Excess Emission}}.
\newblock \emph{\bibinfo{journal}{\apj}} \textbf{\bibinfo{volume}{797}},
  \bibinfo{pages}{85} (\bibinfo{year}{2014}).
\newblock \eprint{1406.6066}.

\bibitem{spitzer_1978}
\bibinfo{author}{{Spitzer}, L.}
\newblock \emph{\bibinfo{title}{{Physical processes in the interstellar
  medium}}} (\bibinfo{year}{1978}).

\bibitem{marx-zimmer_2000}
\bibinfo{author}{{Marx-Zimmer}, M.} \emph{et~al.}
\newblock \bibinfo{title}{{A study of the cool gas in the Large Magellanic
  Cloud. I. Properties of the cool atomic phase - a third H i absorption
  survey}}.
\newblock \emph{\bibinfo{journal}{\aap}} \textbf{\bibinfo{volume}{354}},
  \bibinfo{pages}{787--801} (\bibinfo{year}{2000}).

\bibitem{dickey_2000}
\bibinfo{author}{{Dickey}, J.~M.}, \bibinfo{author}{{Mebold}, U.},
  \bibinfo{author}{{Stanimirovic}, S.} \& \bibinfo{author}{{Staveley-Smith},
  L.}
\newblock \bibinfo{title}{{Cold Atomic Gas in the Small Magellanic Cloud}}.
\newblock \emph{\bibinfo{journal}{\apj}} \textbf{\bibinfo{volume}{536}},
  \bibinfo{pages}{756--772} (\bibinfo{year}{2000}).

\bibitem{lee_2015b}
\bibinfo{author}{{Lee}, M.-Y.}, \bibinfo{author}{{Stanimirovi{\'c}}, S.},
  \bibinfo{author}{{Murray}, C.~E.}, \bibinfo{author}{{Heiles}, C.} \&
  \bibinfo{author}{{Miller}, J.}
\newblock \bibinfo{title}{{Cold and Warm Atomic Gas around the Perseus
  Molecular Cloud. II. The Impact of High Optical Depth on the HI Column
  Density Distribution and Its Implication for the HI-to-H$_{2}$ Transition}}.
\newblock \emph{\bibinfo{journal}{\apj}} \textbf{\bibinfo{volume}{809}},
  \bibinfo{pages}{56} (\bibinfo{year}{2015}).
\newblock \eprint{1504.07405}.

\bibitem{bolatto_2013}
\bibinfo{author}{{Bolatto}, A.~D.}, \bibinfo{author}{{Wolfire}, M.} \&
  \bibinfo{author}{{Leroy}, A.~K.}
\newblock \bibinfo{title}{{The CO-to-H$_{2}$ Conversion Factor}}.
\newblock \emph{\bibinfo{journal}{\araa}} \textbf{\bibinfo{volume}{51}},
  \bibinfo{pages}{207--268} (\bibinfo{year}{2013}).
\newblock \eprint{1301.3498}.

\bibitem{roman-duval_2014}
\bibinfo{author}{{Roman-Duval}, J.} \emph{et~al.}
\newblock \bibinfo{title}{{Dust and Gas in the Magellanic Clouds from the
  HERITAGE Herschel Key Project. II. Gas-to-dust Ratio Variations across
  Interstellar Medium Phases}}.
\newblock \emph{\bibinfo{journal}{\apj}} \textbf{\bibinfo{volume}{797}},
  \bibinfo{pages}{86} (\bibinfo{year}{2014}).
\newblock \eprint{1411.4552}.

\bibitem{paradis_2011}
\bibinfo{author}{{Paradis}, D.} \emph{et~al.}
\newblock \bibinfo{title}{{Spitzer Characterization of Dust in the Ionized
  Medium of the Large Magellanic Cloud}}.
\newblock \emph{\bibinfo{journal}{\apj}} \textbf{\bibinfo{volume}{735}},
  \bibinfo{pages}{6} (\bibinfo{year}{2011}).
\newblock \eprint{1104.1098}.

\bibitem{shaver_1983}
\bibinfo{author}{{Shaver}, P.~A.}, \bibinfo{author}{{McGee}, R.~X.},
  \bibinfo{author}{{Newton}, L.~M.}, \bibinfo{author}{{Danks}, A.~C.} \&
  \bibinfo{author}{{Pottasch}, S.~R.}
\newblock \bibinfo{title}{{The galactic abundance gradient}}.
\newblock \emph{\bibinfo{journal}{\mnras}} \textbf{\bibinfo{volume}{204}},
  \bibinfo{pages}{53--112} (\bibinfo{year}{1983}).

\bibitem{dickinson_2003}
\bibinfo{author}{{Dickinson}, C.}, \bibinfo{author}{{Davies}, R.~D.} \&
  \bibinfo{author}{{Davis}, R.~J.}
\newblock \bibinfo{title}{{Towards a free-free template for CMB foregrounds}}.
\newblock \emph{\bibinfo{journal}{\mnras}} \textbf{\bibinfo{volume}{341}},
  \bibinfo{pages}{369--384} (\bibinfo{year}{2003}).
\newblock \eprint{astro-ph/0302024}.

\bibitem{bertoldi_1992}
\bibinfo{author}{{Bertoldi}, F.} \& \bibinfo{author}{{McKee}, C.~F.}
\newblock \bibinfo{title}{{Pressure-confined clumps in magnetized molecular
  clouds}}.
\newblock \emph{\bibinfo{journal}{\apj}} \textbf{\bibinfo{volume}{395}},
  \bibinfo{pages}{140--157} (\bibinfo{year}{1992}).

\bibitem{kauffmann_2013}
\bibinfo{author}{{Kauffmann}, J.}, \bibinfo{author}{{Pillai}, T.} \&
  \bibinfo{author}{{Goldsmith}, P.~F.}
\newblock \bibinfo{title}{{Low Virial Parameters in Molecular Clouds:
  Implications for High-mass Star Formation and Magnetic Fields}}.
\newblock \emph{\bibinfo{journal}{\apj}} \textbf{\bibinfo{volume}{779}},
  \bibinfo{pages}{185} (\bibinfo{year}{2013}).
\newblock \eprint{1308.5679}.

\bibitem{rosado_1996}
\bibinfo{author}{{Rosado}, M.} \emph{et~al.}
\newblock \bibinfo{title}{{Formation of the nebular complex N11 in the Large
  Magellanic Cloud.}}
\newblock \emph{\bibinfo{journal}{\aap}} \textbf{\bibinfo{volume}{308}},
  \bibinfo{pages}{588--600} (\bibinfo{year}{1996}).

\bibitem{nayak_2016}
\bibinfo{author}{{Nayak}, O.} \emph{et~al.}
\newblock \bibinfo{title}{{Studying Relation Between Star Formation and
  Molecular Clumps on Subparsec Scales in 30 Doradus}}.
\newblock \emph{\bibinfo{journal}{ArXiv e-prints}}  (\bibinfo{year}{2016}).
\newblock \eprint{1608.05451}.

\bibitem{kroupa_2001}
\bibinfo{author}{{Kroupa}, P.}
\newblock \bibinfo{title}{{On the variation of the initial mass function}}.
\newblock \emph{\bibinfo{journal}{\mnras}} \textbf{\bibinfo{volume}{322}},
  \bibinfo{pages}{231--246} (\bibinfo{year}{2001}).
\newblock \eprint{astro-ph/0009005}.

\bibitem{vaidya_2009}
\bibinfo{author}{{Vaidya}, K.}, \bibinfo{author}{{Chu}, Y.-H.},
  \bibinfo{author}{{Gruendl}, R.~A.}, \bibinfo{author}{{Chen}, C.-H.~R.} \&
  \bibinfo{author}{{Looney}, L.~W.}
\newblock \bibinfo{title}{{A Hubble Space Telescope View of the Interstellar
  Environments of Young Stellar Objects in the Large Magellanic Cloud}}.
\newblock \emph{\bibinfo{journal}{\apj}} \textbf{\bibinfo{volume}{707}},
  \bibinfo{pages}{1417--1426} (\bibinfo{year}{2009}).
\newblock \eprint{0910.5901}.

\bibitem{stephens_2017}
\bibinfo{author}{{Stephens}, I.~W.} \emph{et~al.}
\newblock \bibinfo{title}{{Stellar Clusterings around ''Isolated'' Massive YSOs
  in the LMC}}.
\newblock \emph{\bibinfo{journal}{\apj}} \textbf{\bibinfo{volume}{834}},
  \bibinfo{pages}{94} (\bibinfo{year}{2017}).
\newblock \eprint{1609.04399}.

\bibitem{vacca_1996}
\bibinfo{author}{{Vacca}, W.~D.}, \bibinfo{author}{{Garmany}, C.~D.} \&
  \bibinfo{author}{{Shull}, J.~M.}
\newblock \bibinfo{title}{{The Lyman-Continuum Fluxes and Stellar Parameters of
  O and Early B-Type Stars}}.
\newblock \emph{\bibinfo{journal}{\apj}} \textbf{\bibinfo{volume}{460}},
  \bibinfo{pages}{914} (\bibinfo{year}{1996}).

\bibitem{o'dell_1993}
\bibinfo{author}{{O'dell}, C.~R.}, \bibinfo{author}{{Valk}, J.~H.},
  \bibinfo{author}{{Wen}, Z.} \& \bibinfo{author}{{Meyer}, D.~M.}
\newblock \bibinfo{title}{{Identification of velocity systems in the inner
  Orion nebula}}.
\newblock \emph{\bibinfo{journal}{\apj}} \textbf{\bibinfo{volume}{403}},
  \bibinfo{pages}{678--683} (\bibinfo{year}{1993}).

\bibitem{simon-diaz_2003}
\bibinfo{author}{{Sim{\'o}n-D{\'{\i}}az}, S.}, \bibinfo{author}{{Herrero}, A.}
  \& \bibinfo{author}{{Esteban}, C.}
\newblock \bibinfo{title}{{The Trapezium Stars. Preliminary Results on Detailed
  Atmosphere Modeling}}.
\newblock In \bibinfo{editor}{{Reyes-Ruiz}, M.} \&
  \bibinfo{editor}{{V{\'a}zquez-Semadeni}, E.} (eds.)
  \emph{\bibinfo{booktitle}{Revista Mexicana de Astronomia y Astrofisica
  Conference Series}}, vol.~\bibinfo{volume}{18} of
  \emph{\bibinfo{series}{Revista Mexicana de Astronomia y Astrofisica,
  vol.~27}}, \bibinfo{pages}{123--125} (\bibinfo{year}{2003}).

\bibitem{robitaille_2008}
\bibinfo{author}{{Robitaille}, T.~P.}
\newblock \bibinfo{title}{{SED Modeling of Young Massive Stars}}.
\newblock In \bibinfo{editor}{{Beuther}, H.}, \bibinfo{editor}{{Linz}, H.} \&
  \bibinfo{editor}{{Henning}, T.} (eds.) \emph{\bibinfo{booktitle}{Massive Star
  Formation: Observations Confront Theory}}, vol. \bibinfo{volume}{387} of
  \emph{\bibinfo{series}{Astronomical Society of the Pacific Conference
  Series}}, \bibinfo{pages}{290} (\bibinfo{year}{2008}).
\newblock \eprint{0711.4369}.

\bibitem{gaustadt_2001}
\bibinfo{author}{{Gaustad}, J.~E.}, \bibinfo{author}{{McCullough}, P.~R.},
  \bibinfo{author}{{Rosing}, W.} \& \bibinfo{author}{{Van Buren}, D.}
\newblock \bibinfo{title}{{A Robotic Wide-Angle H{$\alpha$} Survey of the
  Southern Sky}}.
\newblock \emph{\bibinfo{journal}{\pasp}} \textbf{\bibinfo{volume}{113}},
  \bibinfo{pages}{1326--1348} (\bibinfo{year}{2001}).
\newblock \eprint{astro-ph/0108518}.

\bibitem{calzetti_2007}
\bibinfo{author}{{Calzetti}, D.} \emph{et~al.}
\newblock \bibinfo{title}{{The Calibration of Mid-Infrared Star Formation Rate
  Indicators}}.
\newblock \emph{\bibinfo{journal}{\apj}} \textbf{\bibinfo{volume}{666}},
  \bibinfo{pages}{870--895} (\bibinfo{year}{2007}).
\newblock \eprint{0705.3377}.

\bibitem{krumholz_2014}
\bibinfo{author}{{Krumholz}, M.~R.} \emph{et~al.}
\newblock \bibinfo{title}{{Star Cluster Formation and Feedback}}.
\newblock \emph{\bibinfo{journal}{Protostars and Planets VI}}
  \bibinfo{pages}{243--266} (\bibinfo{year}{2014}).
\newblock \eprint{1401.2473}.

\bibitem{kruijssen_2014}
\bibinfo{author}{{Kruijssen}, J.~M.~D.} \& \bibinfo{author}{{Longmore}, S.~N.}
\newblock \bibinfo{title}{{An uncertainty principle for star formation - I. Why
  galactic star formation relations break down below a certain spatial scale}}.
\newblock \emph{\bibinfo{journal}{\mnras}} \textbf{\bibinfo{volume}{439}},
  \bibinfo{pages}{3239--3252} (\bibinfo{year}{2014}).
\newblock \eprint{1401.4459}.

\bibitem{kennicutt_1995}
\bibinfo{author}{{Kennicutt}, R.~C., Jr.}, \bibinfo{author}{{Bresolin}, F.},
  \bibinfo{author}{{Bomans}, D.~J.}, \bibinfo{author}{{Bothun}, G.~D.} \&
  \bibinfo{author}{{Thompson}, I.~B.}
\newblock \bibinfo{title}{{Large scale structure of the ionized gas in the
  magellanic clouds}}.
\newblock \emph{\bibinfo{journal}{\aj}} \textbf{\bibinfo{volume}{109}},
  \bibinfo{pages}{594--604} (\bibinfo{year}{1995}).

\bibitem{krumholz_2005}
\bibinfo{author}{{Krumholz}, M.~R.} \& \bibinfo{author}{{McKee}, C.~F.}
\newblock \bibinfo{title}{{A General Theory of Turbulence-regulated Star
  Formation, from Spirals to Ultraluminous Infrared Galaxies}}.
\newblock \emph{\bibinfo{journal}{\apj}} \textbf{\bibinfo{volume}{630}},
  \bibinfo{pages}{250--268} (\bibinfo{year}{2005}).
\newblock \eprint{astro-ph/0505177}.

\bibitem{portegies-zwart_2010}
\bibinfo{author}{{Portegies Zwart}, S.~F.}, \bibinfo{author}{{McMillan},
  S.~L.~W.} \& \bibinfo{author}{{Gieles}, M.}
\newblock \bibinfo{title}{{Young Massive Star Clusters}}.
\newblock \emph{\bibinfo{journal}{\araa}} \textbf{\bibinfo{volume}{48}},
  \bibinfo{pages}{431--493} (\bibinfo{year}{2010}).
\newblock \eprint{1002.1961}.

\bibitem{d'ercole_2008}
\bibinfo{author}{{D'Ercole}, A.}, \bibinfo{author}{{Vesperini}, E.},
  \bibinfo{author}{{D'Antona}, F.}, \bibinfo{author}{{McMillan}, S.~L.~W.} \&
  \bibinfo{author}{{Recchi}, S.}
\newblock \bibinfo{title}{{Formation and dynamical evolution of multiple
  stellar generations in globular clusters}}.
\newblock \emph{\bibinfo{journal}{\mnras}} \textbf{\bibinfo{volume}{391}},
  \bibinfo{pages}{825--843} (\bibinfo{year}{2008}).
\newblock \eprint{0809.1438}.

\bibitem{bekki_2011}
\bibinfo{author}{{Bekki}, K.}
\newblock \bibinfo{title}{{Secondary star formation within massive star
  clusters: origin of multiple stellar populations in globular clusters}}.
\newblock \emph{\bibinfo{journal}{\mnras}} \textbf{\bibinfo{volume}{412}},
  \bibinfo{pages}{2241--2259} (\bibinfo{year}{2011}).
\newblock \eprint{1011.5956}.

\bibitem{schaerer_2011}
\bibinfo{author}{{Schaerer}, D.} \& \bibinfo{author}{{Charbonnel}, C.}
\newblock \bibinfo{title}{{A new perspective on globular clusters, their
  initial mass function and their contribution to the stellar halo and the
  cosmic reionization}}.
\newblock \emph{\bibinfo{journal}{\mnras}} \textbf{\bibinfo{volume}{413}},
  \bibinfo{pages}{2297--2304} (\bibinfo{year}{2011}).
\newblock \eprint{1101.1073}.

\end{thebibliography}

\end{methods}

\end{document}